\documentstyle[prl,aps,multicol,epsf]{revtex}
\begin{document}
\draft
\widetext

\newcommand{\be}{\begin{equation}}
\newcommand{\ee}{\end{equation}}
\newcommand{\ber}{\begin{eqnarray}}
\newcommand{\eer}{\end{eqnarray}}

\title{Generic Transmission Zeros and In-Phase Resonances 
in Time-Reversal Symmetric Single Channel Transport}

\author{H.-W. Lee}
\address{
Center for Theoretical Physics, Seoul National University,
Seoul 151-742, Korea 
}

\maketitle

\widetext
\begin{abstract}
We study phase coherent transport in a single channel system
using the scattering matrix approach.
It is shown that identical vanishing of the transmission amplitude
occurs generically in quasi-1D systems if the time reversal is
a good symmetry. The transmission zeros naturally lead to
abrupt phase changes (without any intrinsic energy scale) and
in-phase resonances, providing insights to recent experiments
on phase coherent transport through a quantum dot. 
\end{abstract}
\pacs{PACS numbers: 73.20.Dx, 73.23.Hk, 73.50.Bk }

\begin{multicols}{2}

\narrowtext
 
In 1995, it was first demonstrated in an experiment
using the Aharonov-Bohm  (AB) interference effect that
the electron transport through a quantum dot contains
a phase coherent component \cite{Yacoby}. This experiment,
however, was found to have some problem due to the so called phase locking
effect \cite{Yeyati}. Two years later, the experiment was
refined using the four probe measurement scheme so that
the phase of the transmission amplitude through the dot 
can be measured in a reliable way \cite{Schuster}. 
It was found that the phase increases by $\pi$ whenever the gate
voltage to the dot sweeps through a resonance 
and that the profile of the phase increase
is well described 
by the Breit-Wigner resonance formula \cite{Breit}.

Unexpected properties were also discovered.
The behavior of the phase evolution is identical (up to $2\pi$)
for a large number of resonances, and between each pair 
of adjacent in-phase resonances there is an abrupt phase change
by $\pi$, whose characteristic energy scale is
much smaller than all other energy scales available in the experiment.
On the other hand, the 1D Friedel sum rule \cite{Harrison}, 
\be
\Delta Q/ e= \Delta \mbox{arg}(t) / \pi  \ ,
\label{eq:1dFriedel}
\ee
predicts that all neighboring resonances are
off phase by $\pi$, which differs from the experimental
findings. Thus two central questions arise: First, 
how can in-phase resonances occur? 
Does it imply that the Friedel sum rule is not
valid for the quantum dot? 
Second, why do abrupt phase changes occur and why are they
so sharp?

Many theoretical investigations addressed these questions. 
It was suggested that the Friedel sum rule is still valid and the
abrupt phase changes are due to ``hidden'' electron charging
events that do not cause conductance peaks \cite{Yeyati}.
It was also speculated that the in-phase resonances are due
to the strong Coulomb interaction \cite{Bruder},
the finite temperature \cite{Oreg}, or the Fano resonance \cite{Xu,Ryu}.
There was also a claim that they are due to peculiar properties
of the AB ring instead of their being a true manifestation of
the phase of the transmission amplitude \cite{Wu}. 
Regarding the characteristic energy scale,
it was claimed that the width of the abrupt phase change is
the true measure of the intrinsic resonance width $\Gamma$
while the measured resonance peaks are thermally broadened
\cite{HackenbroichEuro}. 
 
In this paper, we present a new theory based on
the Friedel sum rule and the time reversal symmetry.
(In Ref.~\cite{Schuster}, the magnetic flux threading the dot
is only a small fraction of a flux quantum.)
One of the key observations is that the 1D Friedel sum 
rule~(\ref{eq:1dFriedel})
is not strictly valid for quasi-1D systems due to
the appearance of the transmission zeros.

\begin{figure}
\epsfxsize=8cm \epsfysize=3.0cm \epsfbox{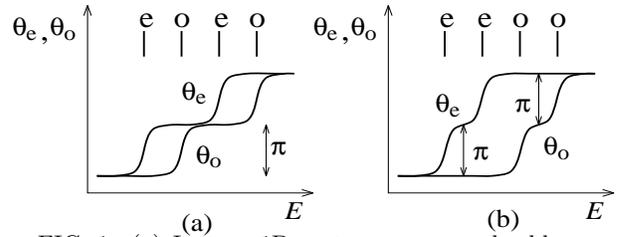}
\caption{(a) In true 1D systems, even and odd resonance levels
alternate in energy. (b) In quasi-1D systems, they do not
necessarily alternate, leading to the transmission zeros
since $t\propto \sin(\theta_{\rm e}-\theta_{\rm o})$.
}
\label{fig:phi}
\end{figure}

To demonstrate this, we first discuss mirror
reflection symmetric systems without magnetic fields.
Since the parity is a good quantum number, 
the scattering states can be decomposed into 
even and odd scattering states: for $|x|>R$,  
$\psi_{\rm e}(x)= e^{-ik|x|}+e^{2i\theta_{\rm e}}e^{ik|x|}$
and 
$\psi_{\rm o}(x)= \mbox{sgn}(x)[e^{-ik|x|}+e^{2i\theta_{\rm o}}e^{ik|x|}]$. 
(For $|x|<R$,  there are scattering potentials which may have higher
dimensional nature as in Ref.~\cite{Schuster}.) 
The outgoing waves are phase-shifted, and the Friedel sum rule 
($\Delta Q_{\rm e} / e=\Delta \theta_{\rm e} / \pi$, 
$\Delta Q_{\rm o} / e=\Delta \theta_{\rm o} / \pi$) 
shows that whenever an even (odd) parity quasibound state
is occupied, $\theta_{\rm e}$ ($\theta_{\rm o}$) shifts by $\pi$
(Fig.~\ref{fig:phi}). 
   
Alternatively, left and right scattering states can be used, 
which are superpositions of
the even and odd scattering states:
$\psi_{\rm l}(x)=[\psi_{\rm e}(x)-\psi_{\rm o}(x)]/2$,   
$\psi_{\rm r}(x)=[\psi_{\rm e}(x)+\psi_{\rm o}(x)]/2$. 
From these relations, one finds that the transmission
amplitude $t$ and $t'$ for the left and right scattering
states are
\be
t=t'=ie^{i\theta}\sin \phi \ ,
\label{eq:tt'}
\ee
where $\theta\equiv \theta_{\rm e}+\theta_{\rm o}$ and
$\phi\equiv \theta_{\rm e}-\theta_{\rm o}$. 
In terms of the new angles, the Friedel sum rule becomes
\be
\Delta Q/e=\Delta \theta/\pi \ .
\label{eq:qtheta}
\ee 

In true 1D systems, even and odd resonant
states alternate in energy and the angle $\phi$
can be limited to the range $0<\phi<\pi$ [Fig.~\ref{fig:phi}(a)].
Then $\Delta \theta=\Delta \mbox{arg}(t)$ and the 1D Friedel
sum rule~(\ref{eq:1dFriedel}) is recovered.

In quasi-1D systems, on the other hand, even and odd levels
do not necessarily alternate in energy. One concrete example
is a dot with the anisotropic harmonic confining potential.
The energy levels of the dot are given by 
$E(n_x,n_y)=\hbar\omega_x (n_x+1/2)+\hbar\omega_y (n_y+1/2)$
where $\omega_x \neq \omega_y$. Here $n_x$ determines
the parity of a level while $n_y$ is a free parameter as
far as the parity is concerned. Because of the presence of
this free parameter, situations like Fig.~\ref{fig:phi}(b)
occur generically, where some of adjacent levels share
the same parity.
Notice that the difference between 
$\theta_{\rm e}$ and $\theta_{\rm o}$ increases
from almost zero to almost $2\pi$ and then decreases
to almost zero. Since the change is continuous, points should
exist where the difference $\phi$ is $\pi$ exactly.
At these points, $\sin \phi$ vanishes identically and
as these points are scanned, $\sin \phi$ reverses its sign,
causing the abrupt phase change of $t$.
It is straightforward to verify that 
the transmission zeros occur whenever neighboring states share the
same parity.

As a result of the transmission zeros,
one finds 
\be
\Delta Q/e=\Delta \theta/\pi \neq \Delta \mbox{arg}(t)/\pi \ .
\ee
Thus the 1D Friedel sum rule~(\ref{eq:1dFriedel})
is not strictly valid for quasi-1D systems.
One immediate consequence is that there are {\it two} possibilities
for adjacent resonances. 
They can be either off phase by $\pi$ or in phase, and
in the latter case, a transmission zero occurs in between.
Another important implication is that there is {\it no}
intrinsic energy scale for the abrupt phase change,
since the transmission zero corresponds to a singular point
as far as the phase is concerned.
It also explains naturally the experimental
observation that the abrupt phase changes occur when
the amplitude of the AB oscillation almost vanishes \cite{Schuster}.

We next generalize the discussion to systems 
without the mirror reflection symmetry.
The electron transport in single channel systems can be described by
the $2\times 2$ scattering matrix ${\bf S}$,
\be
{\bf S}=\left( \begin{array}{cc} \displaystyle
r & t' \\
t & r' 
\end{array} \right)=
e^{i\theta}\left( \begin{array}{cc} \displaystyle
  e^{i\varphi_1}\cos \phi  & i e^{-i\varphi_2} \sin \phi  \\
i e^{i\varphi_2}\sin \phi  &  e^{-i\varphi_1} \cos \phi 
\end{array} \right) \ ,
\label{eq:generalS}
\ee
where the matrix elements are parameterized in a most general
way compatible with ${\bf S^\dagger S=S S^\dagger=I}$.
When the time reversal is a good symmetry, $t=t'$ \cite{Anderson} and 
the angle $\varphi_2$ can be set to zero. 
Then Eq.~(\ref{eq:tt'}) is recovered.
Also the general Friedel sum rule \cite{Langer},
\be
\Delta Q/e=[\Delta \ln \mbox{Det}({\bf S})]/(2\pi i) \ , 
\label{eq:generalFriedel}
\ee
reduces to Eq.~(\ref{eq:qtheta}).
Thus one again finds that both possibilities of
the off-phase resonances and the in-phase resonances
are compatible with the Friedel sum rule and the time reversal
symmetry.

\begin{figure}
\epsfxsize=8cm \epsfysize=3.5cm \epsfbox{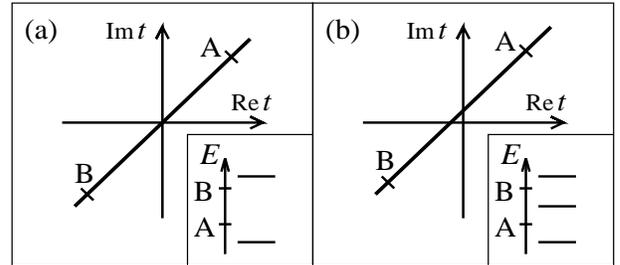}
\caption{Behaviors of the transmission amplitude in the complex
$t$ plane for $\lambda=0$ (mirror symmetric systems) (a) and
$\lambda=\delta \lambda\ll 1$ (nonsymmetric systems) (b).
The behavior in (b) however leads to an unphysical consequence (see text).
Insets show the corresponding energy level diagrams.
}
\label{fig:zero}
\end{figure}

To examine whether the in-phase resonances can appear
generically, one has to investigate whether
the transmission zeros are generic.
The following gedanken experiment is useful for discussion.
Imagine that one changes the confining potential 
$V(x,y;\lambda)=V_{\rm s}(x,y)+\lambda V_{\rm n}(x,y)$ 
of a dot adiabatically by turning on the parameter $\lambda$
where $V_{\rm s}(x,y)=V_{\rm s}(-x,y)$ and 
$V_{\rm n}(x,y)\neq V_{\rm n}(-x,y)$. For $\lambda=0$,
the potential is mirror symmetric and for $\lambda\neq 0$,
the mirror symmetry is broken.
Let us assume that transmission zeros in the mirror symmetric
potential disappear after $\lambda$ is turned on.
Then, Figure~\ref{fig:zero}(a) and \ref{fig:zero}(b) represent the typical 
behaviors of the transmission amplitude in the complex $t$ plane
for $\lambda=0$ and $\lambda=\delta \lambda \ll 1$, respectively.
Notice that there is no transmission zero in \ref{fig:zero}(b)
since the trajectory of $t$ is shifted off the origin.
As the energy is scanned from $A$ to $B$, $\Delta \theta=0$
in Fig.~\ref{fig:zero}(a). 
In Fig.~\ref{fig:zero}(b), on the other hand,
$\Delta \theta=\pi$ and thus $\Delta Q=e$. 
The corresponding energy levels of the dot
are depicted in the insets. 
While there is no energy level between $A$ and $B$
for $\lambda=0$, 
a new level is present between $A$ and $B$ in the level diagram
for $\lambda=\delta \lambda$ since $\Delta Q=e$. 
Upon the infinitesimal change of the confining potential, however,
new energy levels cannot appear suddenly 
although they can drift up and down. 
Thus this sudden appearance of a new energy level is unphysical
and to avoid this, 
the trajectory for $\lambda=\delta \lambda$
should pass through the origin.
This argument applies all along the turning on process and 
it shows that the transmission
zeros should still appear generically even if the system is not
mirror symmetric.

\begin{figure}
\epsfxsize=8cm \epsfysize=3cm \epsfbox{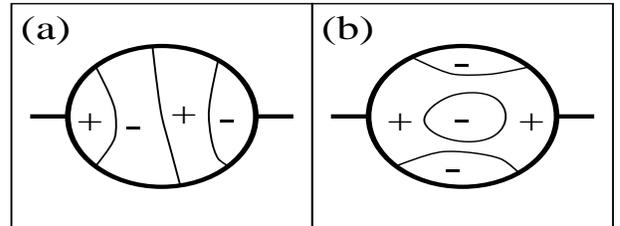}
\caption{Two classes of wave function nodes (thin solid line)
in two dimensions. The marks $(+)$ and $(-)$ 
represent the fact that the wave functions in the marked areas 
have positive and negative values, respectively.
Notice that while each spanning node in (a) shifts
the resonance phase by $\pi$, the introduction of 
nonspanning nodes (b)
does not affect the resonance phase.
Straight lines on the left and right represent 1D electrodes.
}
\label{fig:2dnodes}
\end{figure}

One can also argue for the in-phase resonances
directly, which then establishes the appearance of the transmission
zeros since these two features are linked to each other.
With the time-reversal symmetry, the wave functions can be taken
as real. In true 1D systems, the number of wave function nodes
increases by 1 when a new level appears 
(oscillation theorem \cite{Landau}), and each node
increases the phase of the transmission amplitude by $\pi$.
In quasi-1D systems, on the other hand, 
there are two classes of nodes:
``spanning'' nodes [Fig.~\ref{fig:2dnodes}(a)] that
connect two opposite boundaries of the dot, and
``nonspanning'' nodes [Fig.~\ref{fig:2dnodes}(b)]
that touch either only one particular boundary or no boundary at all.
Such nonspanning nodes can be created, for example, by 
excitations in the transverse direction
or by negative impurity potentials in the interior of the dot.
The two classes of nodes affect
the phase of the transmission amplitude in different ways.
While each spanning node shifts the phase by $\pi$,
nonspanning nodes do {\it not} affect the phase at all.
In the experiment \cite{Schuster}, the transverse size of 
the quantum dot is estimated to be much larger than 
the Fermi length. 
In such a case, nonspanning nodes are equally plausible
as spanning nodes, and accordingly in-phase
resonances can occur as 
generically as off-phase resonances. 

Until now, we have demonstrated the generic appearance of
the transmission zeros and in-phase resonances based on
the Friedel sum rule and the time-reversal symmetry.
Next we demonstrate that multiple
resonances also lead to the transmission zeros naturally
if the time reversal is a good symmetry. 
(This demonstration, in fact, constitutes an alternative
derivation of the same conclusion without using the Friedel sum rule.) 
Near a resonance, the scattering matrix 
becomes ${\bf S}(E)={\bf S}_{\rm bg}-i{\bf B}_0/(E-E_0+i\Gamma_0/2)$
where the $2\times 2$ matrix ${\bf S}_{\rm bg}$ 
(${\bf S}_{\rm bg}^\dagger{\bf S}_{\rm bg}={\bf I}$) is
the energy independent background contribution \cite{Taylor}.
If the off-diagonal matrix elements of ${\bf S}_{\rm bg}$ are
sufficiently small, ${\bf S}(E)$ describes the Breit-Wigner resonance.
For multiple resonances, the scattering matrix becomes
\be
{\bf S}(E)= {\bf S}_{\rm bg} - \sum_{k=1}^N 
  {i{\bf B}_k \over E-E_k+i\Gamma_k/2} \ .
\label{eq:poles1}
\ee
Here we emphasize that the matrix residues $-i{\bf B}_k$ are not independent.
Instead they should be highly correlated so that
${\bf S}^\dagger (E){\bf S}(E)={\bf I}$ for arbitrary real $E$.
(This is the origin of the limited
phase relations between resonances.)
From the unitarity relation and 
the time-reversal symmetry, one
finds five constraints :
$|t(E)|^2=|t'(E)|^2$, $|r(E)|^2=|r'(E)|^2$;  
$|t(E)|^2+|r(E)|^2=1$; 
$t(E)/t' (E)^*+r(E)/r'(E)^*=0$; 
$t(E)=t'(E)$. 

It turns out that to examine the implications of the constraints,
it is more convenient to express the matrix
elements of ${\bf S}(E)$ in the product representation
by summing up all $N+1$ terms in Eq.~(\ref{eq:poles1}).
\ber
t(E)=t'(E)&=&t_{bg}\prod_{k=1}^{N}
  {E-\varepsilon_k +i\mu_k/2 \over E-E_k+i\Gamma_k/2} \ ,
  \nonumber \\
r(E)&=& r_{bg} \prod_{k=1}^{N}
  {E-\epsilon_k+i\nu_k/2 \over E-E_k+i\Gamma_k/2} \ ,
  \label{eq:elements1} \\
r'(E)&=& r'_{bg} \prod_{k=1}^{N}
  {E-\epsilon_k+i\nu'_k/2 \over E-E_k+i\Gamma_k/2} \ ,
  \nonumber
\eer
where $(\nu_k)^2=(\nu'_k)^2$ \cite{zeronumber}.
Then by imposing the constraints and the condition of
no degenerate resonance levels \cite{degeneracy}, one finds
\be 
\mu_k=0 \mbox{ for all } k \ ,
\ee
which implies that all zeros of $t(E)$ are located on
the real energy axis.
(We mention that impurities and irregular boundaries of the dot
generate the level repulsion that lifts the degeneracy. 
The degeneracy is lifted further by the Coulomb blockade effect.)

\begin{figure}
\epsfxsize=8cm \epsfysize=4.5cm \epsfbox{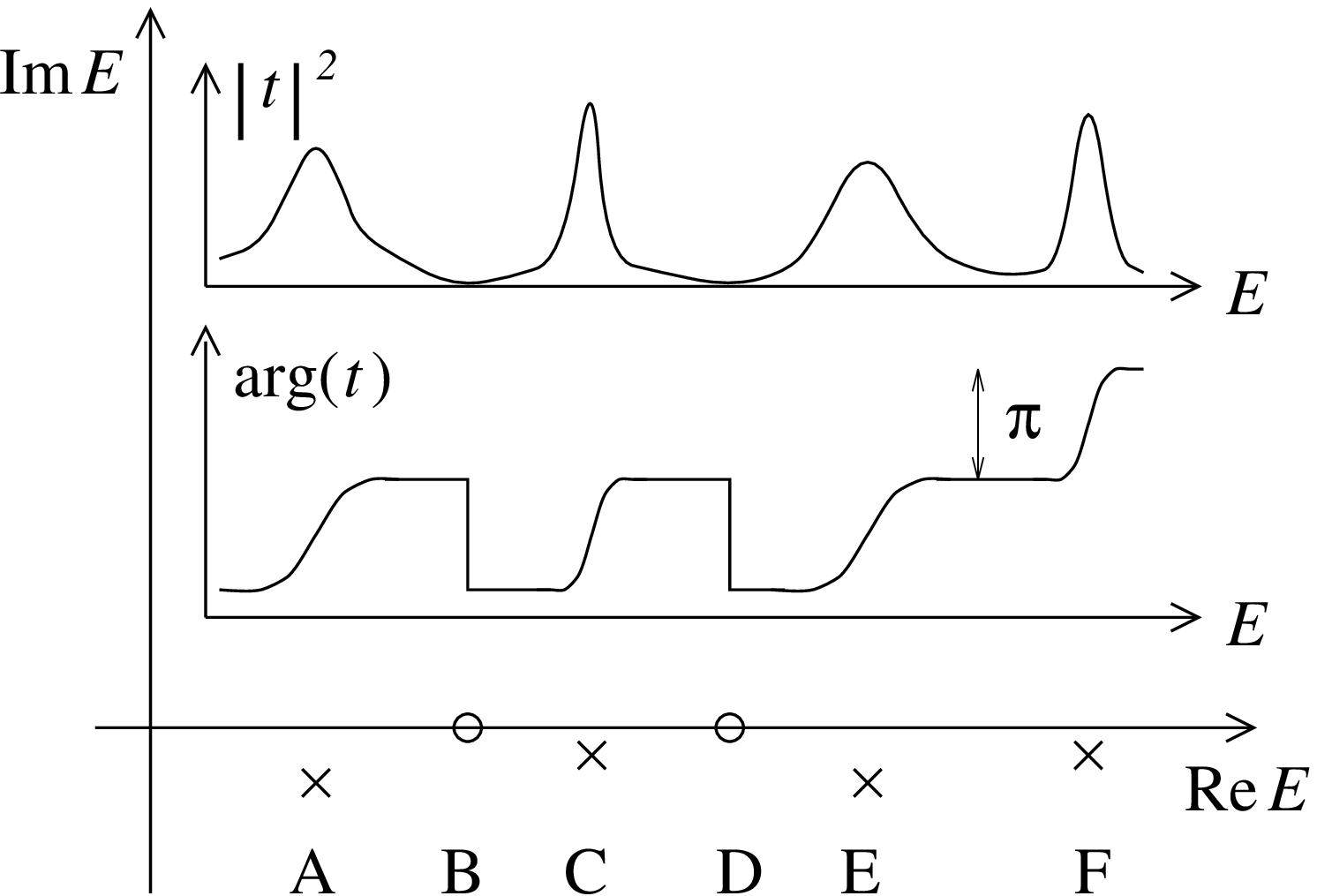}
\caption{Zeros $(\circ)$ and poles $(\times)$ of the transmission
amplitude $t(E)$ in the complex energy plane. 
Insets show the corresponding behaviors of the magnitude
$|t|^2$ and the phase $\mbox{arg}(t)$ as a function
of real energy $E$.
}
\label{fig:zero_pole}
\end{figure}

Evolution of the transmission amplitude is determined from
the locations of poles and zeros. Thus this analysis produces
the following prediction (Fig.~\ref{fig:zero_pole});
If there is a transmission zero ($B$ and $D$) in between,
two neighboring resonances ($A$-$C$ and $C$-$E$) are in phase,
and otherwise, they are off phase ($E$-$F$).
Notice that these predictions are identical to those of
the Friedel sum rule.

It is instructive to compare the transmission zeros from
the Breit-Wigner resonances and the Fano resonances \cite{Xu,Ryu}.
Within an energy window that contains two Breit-Wigner resonances,
$t(E)=-i({\bf B}_k)_{21}/(E-E_k+i\Gamma_k/2)
-i({\bf B}_{k+1})_{21}/(E-E_{k+1}+i\Gamma_{k+1}/2)$. 
By summing up the two contributions, one finds 
$t(E)=\alpha (E-\beta)/[(E-E_k+i\Gamma_k/2)(E-E_{k+1}+i\Gamma_{k+1}/2)]$
where $\beta=\beta^*$ due to the time-reversal symmetry.
Thus the transmission zero at $E=\beta$ is due to
the completely ``destructive interference'' 
of the {\it two} resonance levels.

The transmission zeros also arise from the Fano
resonance \cite{Xu,Ryu}, to which the same 
expression~(\ref{eq:poles1}) applies.
Unlike the Breit-Wigner resonances, however,
the off-diagonal matrix elements of ${\bf S}_{\rm bg}$ are
not small. Thus near a Fano resonance, one finds
$t(E)=({\bf S}_{\rm bg})_{21}-i({\bf B}_k)_{21}/(E-E_k+i\Gamma_k/2)
=\alpha(E-\beta)/(E-E_k+i\Gamma_k/2)$ where
$\beta=\beta^*\approx E_k$.
One finds again the transmission zero. 
It should be noted however that the transmission zero
is now due to the destructive interference
of the background contribution (continuous state of the energy channel) 
and the pole contribution (localized state
for Ref.~\cite{Xu} and $t$-stub for Ref.~\cite{Ryu}).
Notice also that the Fano resonance peak is highly asymmetric 
since $\beta\approx E_k$, which disagrees with 
the experiment \cite{Schuster}.

Below we discuss briefly effects of the electron-electron interaction
and magnetic fields on the transmission zeros. 
Langer and Ambegaokar \cite{Langer}
have shown that even in the presence of the interaction,
the general Friedel sum rule~(\ref{eq:generalFriedel})
is valid at $E=E_F$ provided that quasiparticle excitations
at the Fermi energy remain well defined. Then all analyses
for noninteracting systems apply equally to interacting
systems 
if one fixes the probing energy $E$ at $E_F$ and instead varies
the depth of the potential well,
which amounts to replacing
$E$ by $E_F-\eta eV_g$. 
(Fig.~\ref{fig:zero} can be used to argue against the
disappearance of the transmission zeros upon the adiabatic
interaction turning on. In this case, $\Delta Q=e$ 
in Fig.~\ref{fig:zero}(b) can
be interpreted as the sudden charge density jump.)

Magnetic fields, on the other hand, affect the transmission zeros
in a fundamental way since it breaks the time-reversal symmetry.
In this case,  $t\neq t'$ and 
the angle $\varphi_2$ in Eq.~(\ref{eq:generalS})
can have nonzero values.
Then the transmission zeros are generically replaced by
the rapid but continuous change of $\varphi_2$ by $\pi$,
and thus a finite energy scale appears for the
abrupt phase changes. The precise energy scale
depends on the detailed electron dynamics inside
the dot, which goes beyond the scope of this paper.

Lastly, we discuss the large dominance of the in-phase resonances 
over the off-phase resonances in Ref.~\cite{Schuster}.
Hackenbroich {\it et al.} \cite{Hackenbroich} proposed that
avoided crossings of single particle levels may result
in a long sequence of  resonances carrying the
{\it same} internal wave function. 
In view of the present analysis, this mechanism is
overrestrictive since it restricts not only
the number of spanning nodes but also
the number of nonspanning nodes as well.
We speculate that a less restrictive and possibly
more widely applicable mechanism may exist
which exploits the ``degree of freedom'' given
by the nonspanning nodes.
Further investigation in this direction is necessary.

In summary, we demonstrated that the transmission zeros 
and the in-phase resonances are
generic features in time-reversal symmetric
single channel transport if the transverse size
of a scatterer (dot) is sufficiently larger than
the Fermi wavelength.

The author acknowledges M. Y. Choi for helpful suggestions at
the initial stage of this work, 
G. S. Jeon for helpful discussions,
and M. B\"uttiker for critical comments on the manuscript.
He also thanks C.-M. Ryu for providing his paper before publication.
This work is supported by the Korea Science and Engineering Foundation
through the SRC program at SNU-CTP and through the KOSEF  
fellowship program.

\end{multicols}


\begin{references}
\bibitem{degeneracy} The constraints alone allow nonzero
  $\mu_k$ provided that  ${\bf S}(E)$ itself includes the factor
  $(E-\varepsilon_k+i\mu_k/2)/(E-\varepsilon_k-i\mu_k/2)$
  that is of magnitude 1 for arbitrary real $E$.
  This factor is equal to ${\bf I}-i{\bf A}_k/(E-\varepsilon_k-i\mu_k/2)$
  where ${\bf A}_k=-\mu_k {\bf I}$.
  One then finds that the pole $\varepsilon_k+i\mu_k/2$ represents
  doubly degenerate resonance levels since 
  the rank of the residue matrix is equal to 
  the degeneracy \protect\cite{Taylor} and the rank of
  ${\bf A}_k$ is two.
\end{references}
\end{document}